\documentclass[aps,pre,showpacs,reprint,superscriptaddress]{revtex4-2}
\usepackage{graphicx,color,epsfig}
\usepackage{amssymb,amsmath,dcolumn,bm}

\begin{document}
\title{Structural transitions induced by adaptive rewiring in networks with fixed states}
\author{R. C\'ardenas-Sabando} \email{Current address: Max Planck Institute for the Physics of Complex Systems, Dresden, 01187, Germany}
\affiliation{School of Physical Sciences \& Nanotechnology, Universidad Yachay Tech, 100119 Urcuqu\'i, Ecuador.}
\author{M. G. Cosenza}
\affiliation{School of Physical Sciences \& Nanotechnology, Universidad Yachay Tech, 100119 Urcuqu\'i, Ecuador.}
 \author{J. C. Gonz\'alez-Avella}
\affiliation{Institute for Cross-Disciplinary Physics and Complex Systems, UIB-CSIC, Palma de Mallorca, Spain.}
\date{Chaos, Solitons, \& Fractals \textbf{206}, 117954 (2026)}

\begin{abstract}
We investigate structural transitions in adaptive networks where node states remain fixed and only the connections evolve via state-dependent rewiring. Using a general framework characterized by probabilistic rules for disconnection and reconnection based on node similarity, we systematically explore how homophilic and heterophilic interactions influence network topology. A mean-field approximation for the stationary density of active links—those connecting nodes in different states—is developed to determine the conditions under which fragmentation occurs. Analytical results closely agree with numerical simulations. To distinguish community formation from fragmentation, we introduce order parameters that integrate modularity and connectivity. This enables the characterization of three distinct network phases on the rewiring parameter space: i) random connectivity, ii) community structure, and iii) fragmentation. Community structure emerges only under moderate homophily, while extreme homophily or heterophily lead to fragmentation or random networks, respectively. These findings demonstrate that adaptive rewiring alone, independent of node dynamics, can drive complex structural self-organization, with implications for social, technological, and ecological systems where node attributes are intrinsically stable.
\end{abstract}


\maketitle

\section{Introduction}
Many systems in physics, sociology, biology, and technology can be described as
adaptive or coevolutionary networks, in which the connections and the states of the nodes influence each other and evolve simultaneously~\cite{Zimmermann,Bornholdt,Ito,Egui,Gross,Kozma,Risau,Sayama}.
Coevolutionary dynamics
gives rise to rich phenomena such as network fragmentation \cite{Holme,Fede,Bhome}, community formation \cite{Kaski,Mandra,Tucci}, synchronization \cite{Ott,Boccaletti}, and polarization \cite{Maia,Li}. Consequently, this topic has attracted much interest in recent years.

Models of coevolutionary systems generally consider the coexistence of
 two processes: the dynamics of states of the nodes
and the rewiring of the network connections.
The collective behaviors emerging in coevolutionary systems depend
on the competition between the time scales of these two coexisting processes.
However, in some systems, the states of the nodes remain static and only
the network structure evolves in response to those states. This less studied situation
corresponds to the limit where the difference between these two time scales is pronounced, with the network topology changing more rapidly than the node states.

In this article we investigate a general framework for adaptive rewiring dynamics in networks with fixed node states. Unlike models where node states coevolve with links, we analyze purely topological evolution governed by state-dependent disconnection and connection probabilities. By isolating the
 effects of adaptive rewiring,  we focus on the investigation of how homophilic or heterophilic interactions affect the phenomena of fragmentation and community formation on networks.
This framework is particularly relevant to systems where agent characteristics are stable over time, but connections adapt. For instance, in societies with entrenched ideological or identity-based positions, individuals may maintain their opinions or cultural traits, while social links form or dissolve based on similarity or dissimilarity.  Such dynamics are observed in residential segregation, where individuals of fixed identities relocate based on neighborhood composition \cite{Schelling}, or in collaborative or institutional networks, where researchers often maintain fixed disciplinary identities, while partnerships adapt to contextual or funding-driven criteria \cite{Newman}.  In technological networks, such as Internet of Things systems, many device specifications are static and links adapt based on compatibility criteria \cite{Zanella}. In ecological systems, species with fixed functional roles (e.g., predator or pollinator) may change their interactions due to spatial or environmental factors \cite{Bascompte}.
An analogous fixed-state scenario appears in the  continuous swarmalator model, in the limit
when there is no direct phase interaction between agents but they can move in space \cite{Okeefe,Yadav}.

In Section II we present the adaptive rewiring model with fixed states.
We employ a general framework in which the rewiring process is expressed in terms of disconnection and reconnection probabilities represented by two parameters describing the likelihood of breaking or forming links between nodes with similar states \cite{Herrera2011}. This parametrization allows us to explore, in a unified way, a continuum transition between purely homophilic and purely heterophilic regimes.
Section III contains the mean-field approach and the analytical solution for the stationary density of active links -- those connecting nodes in different states. We compare the analytic solution with numerical simulations and find close agreements. Section IV describes the formation of communities and fragmentation in a network.
While it is intuitively expected that extreme homophily leads to fragmentation, it is not obvious under what conditions a community structure can be sustained without breaking the network apart.
We show that the formation of communities is related to a decrease in the density of active links.
To distinguish community or modular structure from mere fragmentation, we introduce combined topological order parameters that integrate measures of modularity and connectivity. This allows us to characterize three distinct configurations or phases on the space of parameters controlling the rewiring process: I) a randomly connected network; II) community structure, and  III)  a fragmented network.
We find that moderate levels of homophily are necessary for the formation of communities, whereas extreme homophily or heterophily  lead instead to network segregation or random connectivity.
Section V presents our conclusions.

\section{Adaptive rewiring in networks}
Consider an initial undirected random network of $N$ nodes having average degree  $\bar{k}$, i.e., $\bar{k}$ is the average number of neighbors for each node.
Let $\nu_i$ be the set of neighbors of node $i$ at a given time, $i=1,\ldots,N$. We denote by $g_i$ the state variable of node $i$ which we assume to be discrete; i.e., $g_i$
can take any of $G<N$ possible values.
The states $g_i$ are initially assigned at random with a uniform distribution.
We assume that the states of the nodes do not change; that is, there is no node dynamics. Only the links between nodes may evolve through a process of adaptive rewiring that takes into account the states of the nodes.
Since
the states of the nodes are fixed,
the average number of nodes in each state in the network will remain  at $N/G$.

In the general framework for coevolutionary dynamics in networks introduced in \cite{Herrera2011}, any rewiring process, independently of the rate at which it is exerted,  can be described in terms of
two basic actions: disconnection and connection between nodes.
  Either action of disconnection or connection
 depends on some criteria for
   comparison of the states of
  the nodes that can be represented by two parameters.
In this article, we employ
the parameter $d \in [0, 1]$ that represents the probability that two nodes in
identical states become disconnected, so that $1 - d$ is the probability that two nodes in different states disconnect
from each other. Similarly, the parameter $r \in [0, 1]$ denotes the probability that two nodes in identical states become
connected, and $1 - r$ is the probability that two nodes in different states connect to each other. Then, a specific
adaptive rewiring process can be characterized by a pair
of values $d$ and $r$,
corresponding to
a point on the space of parameters $(d,r)$.
For example, a random rewiring process with disconnection and reconnection actions that do not take into account the states of the nodes, as in the Watts-Strogatz model of small-worlds \cite{Watts}, corresponds to the point $(0.5, 0.5)$,

We define
the dynamics of the network topology under an adaptive rewiring
process $(d,r)$
by the following iterative algorithm:
\begin{enumerate}
\item Choose randomly a node $i$ such that $k_i>0$.
\item Break the edge between $i$ and a neighbor $j \in \nu_i$ with probability $d$
if $g_i=g_j$, and with probability $(1-d)$ if $g_i\neq g_j$.
\item Connect node $i$ with a node $l \notin \nu_i$ with probability $r$ if
$g_i=g_l$, and with probability $(1-r)$ if $g_i\neq g_l$
\end{enumerate}

This rewiring process
guarantees the conservation of the total number of links in the network.
A time step corresponds to $N$ iterates of this algorithm.
We have verified that the collective behavior of this system is statistically invariant if steps 2 and 3 are interchanged.

\section{Analytic mean field solution for the density of active links}
The average number of
links in the network is $(\bar k N)/2$. A link is called \textit{active} if it connects nodes in different states, while a link connecting nodes in the same state is said \textit{inert}.
Let $n(t)$ be the number of active links in the network at time t.
Then the density, or average fraction, of active links in the network at a given time is $\rho(t)=2n(t)/(\bar k N)$, and the average fraction of inert links is $1-\rho(t)$.

We consider a mean field approximation where the system is assumed to be
homogeneous. Then, the probability that a given link is active can be estimated by the average density of active links $\rho$.
We calculate the change
in the number of active links in one time step as follows.
The probability that an inert link between node $i$ and node $j  \in \nu_i$ is randomly
selected and becomes active after one update is $d(1-r)$. Since the average fraction of inert links in the network is $(1-\rho)$, then the
average increase in the number of active links in one update is $\Delta n_+=d(1-r)(1-\rho)$. Similarly,
the probability that an active link between $i$ and $j \in \nu_i$ is randomly
selected and becomes inert after one update is $r(1-d)$. Then, since $\rho$ is the average fraction of active links in the network, the
average decrease in the number of active links in a time step is $\Delta n_-=r(1-d)\rho$. Thus, the change in the number of active links in one update  will be
\begin{equation}
\label{Delta}
\Delta n= \Delta n_+-\Delta n_- =  d(1-r)(1-\rho)-r(1-d) \rho .
\end{equation}

Since $n= \frac{\bar k N}{2} \rho$, the change in the number of active links is $\Delta n= \frac{\bar k N}{2} \Delta \rho$. This yields
\begin{equation}
\label{Drho}
\Delta \rho = \frac{2}{\bar k N}  [d(1-r)(1-\rho)-r(1-d) \rho] .
\end{equation}
Thus, the change in the average density of active links over a time  interval $\Delta t = \frac{1}{N}$, in the limit $N \to \infty$,
can be expressed by the following differential equation,
\begin{equation}
 \label{Deq}
 \frac{d\rho}{dt} = \frac{2}{\bar k}  \left[ d(1-r)(1-\rho)-r(1-d) \rho\right] .
\end{equation}

Equation (\ref{Deq}) can be seen as a balance relation between
creation and disappearance of active links, where the first term on the right side of the equation describes an increase in the
number of active links, while the second term expresses a decrease of this number.

Equation~(\ref{Deq}) can be written as
\begin{equation}
 \label{Deq2}
 \frac{d\rho}{dt} = \frac{2}{\bar k} \left[ d(1-r)-(d+r-2rd)\rho\right] ,
\end{equation}
and its solution is
\begin{equation}
\label{sol}
\rho(t)= \rho^*-(\rho^*-\rho_0) e^{-\frac{2}{\bar k}(d+r-2rd)t} ,
\end{equation}
where $\rho_0$ is the initial density of active links in the network, and
\begin{equation}
\rho^*= \frac{(1-r)d}{r(1-2d)+d}
\label{rhos}
\end{equation}
is the asymptotic or stationary solution of Eq.~(\ref{Deq2}).

The curves of constant stationary density $\rho^*=C=\mbox{constant}$, are given by
\begin{equation}
d= \frac{Cr}{(1-r)+C(2r-1)} .
\label{crho}
\end{equation}

\begin{figure}[h]
    \centering
    \includegraphics[scale=0.6]{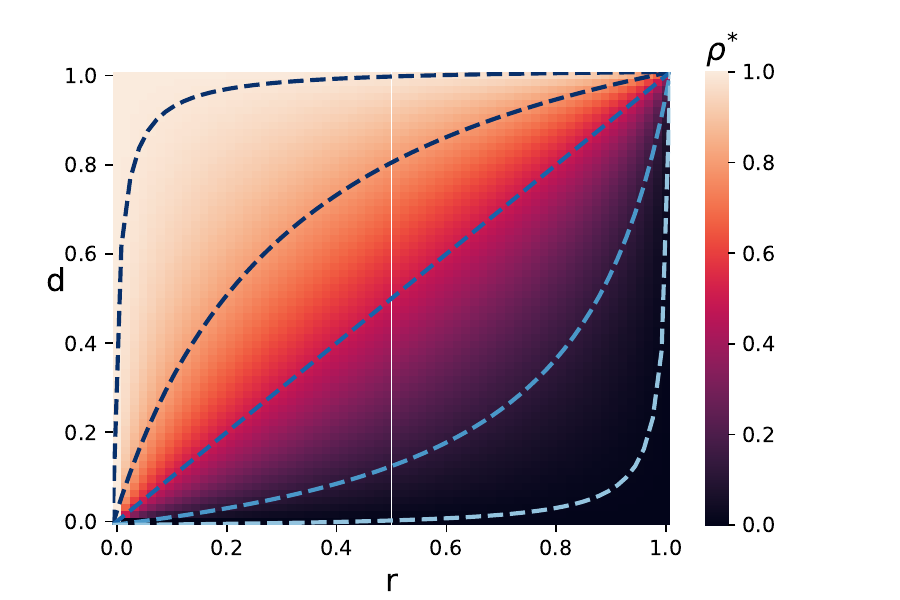}
    \caption{Stationary density of active links $\rho^*$, given by the analytic solution Eq.~(\ref{rhos}), on the space of parameters  $(r,d)$. The values of $\rho^*$ are represented by a color code indicated by a bar on the right.  Darker colors correspond to small values of $\rho^*$, while lighter colors represent higher values of $\rho^*$. Curves for constant values of $\rho^*$, Eq.~(\ref{crho}), are plotted with dashed lines. The boundary lines  $d=0$ and $r=1$, where $\rho^*=0$, correspond to fragmentation of the network.}
    \label{fig:f2}
\end{figure}

Figure~\ref{fig:f2} shows the stationary density of active links $\rho^*$
 on the plane the parameters  $(d,r)$
given by Eq.~(\ref{rhos}).
For each point $(d, r)$ on this plane, the value
$\rho^*$ is displayed by means of a color code indicated in
 the figure. Curves of constant $\rho^*$ values, described by Eq.~(\ref{crho}), are also shown.
 Higher densities of active links occur for parameter values $d \to 1$, $r \to 0$, while lower densities of this quantity correspond to the values
$d \to 0$, $r \to 1$.

Note that if $\rho_0 > \rho^*$ in Eq.~(\ref{sol}), the density $\rho(t)$ decays in time to the stationary value $\rho^*$. Conversely,
if  $\rho_0 < \rho^*$, the density $\rho(t)$ increases up to the value $\rho^*$.

\begin{figure}[h]
    \centering
    \includegraphics[scale=0.54]{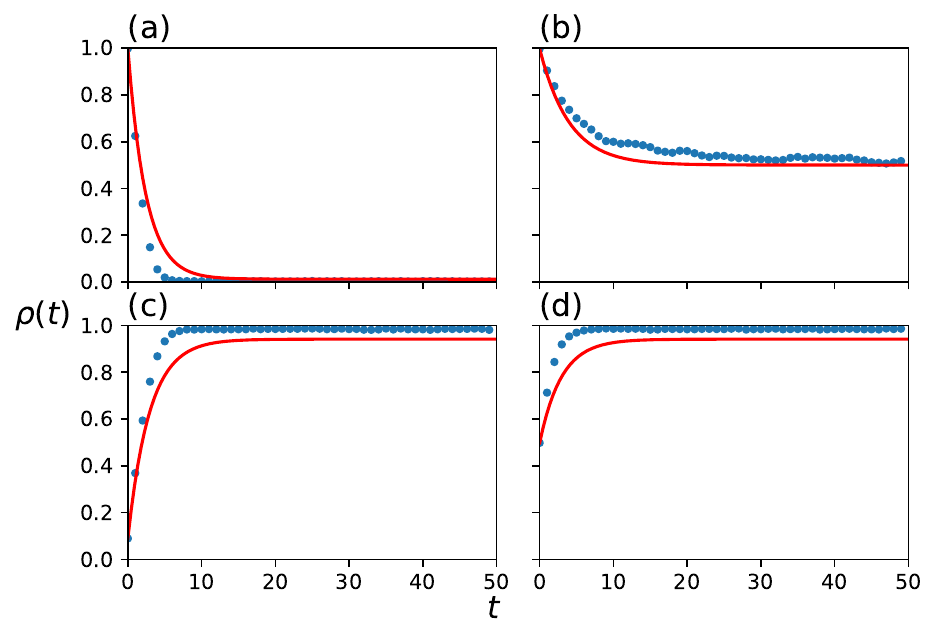}
    \caption{Evolution of the density of active links as a function of time for different rewiring processes $(r,d)$ on initial random networks. The continuous red line is the analytic solution $\rho(t)$ in Eq.~(\ref{sol}) and the blue dots correspond to the density of active links calculated from numerical simulations. The simulations were performed on random initial networks with parameters $N=3200$, $\langle k \rangle=4$ and $G=320$.
    (a) $(r,d)=(0.9,0.1)$. (b) $(r,d)=(0.5,0.5)$. (c) $(r,d)=(0.2,0.8)$. (d) $(r,d)=(0.2,0.8)$, different initial density of active links.}
    \label{fig:f1}
\end{figure}

Figure~\ref{fig:f1} shows the evolution of the density of active links in the network obtained from both, the analytical solution Eq.~(\ref{sol}) and numerical simulations based on the algorithm, for different values of parameters $(r,d)$ and different initial random conditions. The simulations are in good agreement with Eq.~(\ref{sol}); approaching the stationary  solution $\rho^*$ in relatively short times. In this article, we keep the ratio $N/G=10$ for all simulations.

\begin{figure}[h]
    \centering
    \includegraphics[scale=0.55]{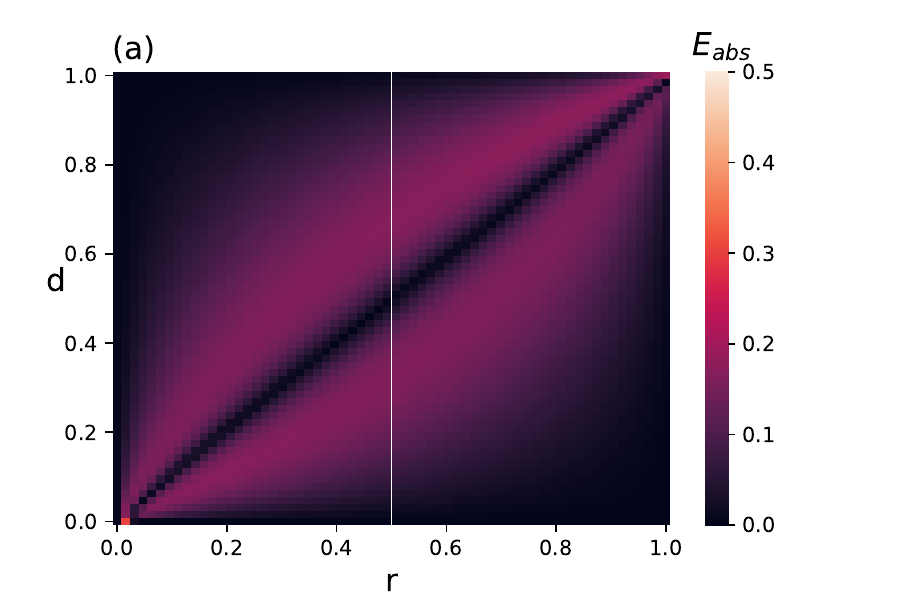}
   \includegraphics[scale=0.5]{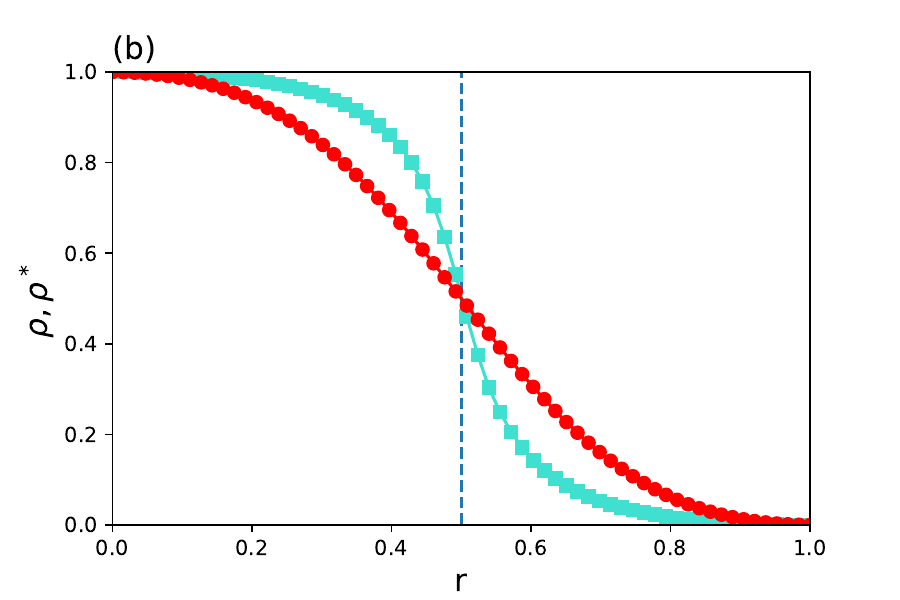}
\caption{(a) Error  $E_{\mbox{\tiny{abs}}}=|\rho^* - \rho|$  between the analytic solution $\rho^*$ and the asymptotic value of the density
of active links $\rho$ calculated from numerical simulations, plotted on the plane $(r,d)$. (b) Comparison of the analytic stationary solution
 $\rho^*$ (red circles) and the numerical simulation asymptotic value $\rho$ (blue squares) as functions of
$r$ along the diagonal line $d = 1 - r$.  Network
parameters for the simulations are $N = 3200$, $G = 320$, $\langle k \rangle = 4$. Each data point shown in the simulation corresponds to the average over $100$ realizations of initial conditions on the network.}
    \label{fc}
\end{figure}

The  mean field solution assumes infinite size and complete homogeneity of the
density of active links for all values of parameters $(d, r)$. These simplifying assumptions are not achieved in the numerical
simulations.
Thus, one should not expect complete agreement between the mean
field model and the simulations for all parameter values.
Figure~\ref{fc}(a) shows the deviation $E_{\mbox{\tiny{abs}}}=|\rho^* - \rho|$  between the analytic stationary solution
 $\rho^*$ and the asymptotic value of the density
of active links $\rho$ calculated from numerical simulations.
The agreement between the stationary solution and the simulations is within $10\%$.
In Fig.~\ref{fc}(b) the quantities  $\rho^*$ and $\rho$ are plotted as functions of $r$  for values of $d$ along the diagonal line $d = 1 - r$.

The parameter regime $d \to 1$, $r \to 0$ corresponds to strong heterophily, where connections between nodes in different states are strongly favored. In this limit, one expects a high density of active links ($\rho \to 1$), consistent with the simulation results. Conversely, the regime $d \to 0$, $r \to 1$ represents strong homophily, where connections between nodes in the same state are highly probable. This leads to a pronounced reduction in the number of active links ($\rho \to 0$), again in agreement with the simulations. The mean-field approximation predicts $\rho^* = 1$ in the heterophilic limit and $\rho^* = 0$ in the homophilic limit, showing good agreement with the numerical simulations in these regions, as seen in Fig.~\ref{fc}(b)).
 In both extremes, the network generated by the simulation approaches a more homogeneous distribution of active links, as assumed in the mean-field description.
Along the line $d = r$, the probabilities of disconnection and connection between nodes in the same or in different states are equal. In this symmetric regime, one may expect that, on average,  half of the links in the network remain active. The mean-field analysis gives $\rho^* = 0.5$, and the numerical simulation yields values very close to $\rho = 0.5$ throughout the line $d = r$. Consequently, the absolute error, $E_{\text{\tiny abs}} = |\rho^* - \rho|$, remains small along this line, as illustrated in Fig.~\ref{fc}(a).
Outside these limiting cases, however, the simulations exhibit larger deviations from the mean-field model.
These discrepancies arise due to correlations and finite size effects appearing in
the rewiring dynamics of the simulations. The numerical simulations do not
 generate networks with perfectly homogeneous distributions of active links as assumed in the  mean-field analysis.

\section{Fragmentation and formation of communities}

In the mean field limit $N \to \infty$, the solution Eq.~(\ref{rhos}) indicates that the stationary density of active links
 $\rho^*$ becomes zero for $d=0$ or $r=1$. Rewiring processes with parameter values $d=0$ or $r=1$
 completely favor the connection between nodes in similar states,
and therefore reduce the number of active links in the network until their density vanishes.
Then, the condition $\rho^*=0$  means that the network consists of several components or subgraphs disconnected from each other, with all nodes in a subgraph sharing the same state, with different states for each subgraph -- a phenomenon referred to as network fragmentation.
Thus, the boundary lines $d=0$ and $r=1$ on the plane $(d,r)$ in Fig.~\ref{fig:f2} correspond to a fragmented network.
Actually, coevolutionary systems exhibiting network fragmentation reported in the literature used rewiring processes that can be represented as points on these lines. For example,
 a rewiring process $(d=0.5, r=1)$ was employed in Ref.~\cite{Holme}; $(d=0, r=1)$ characterizes that used in Ref.~\cite{Fede};  $(d=1, r=1)$ can be identified in Ref.~\cite{Bhome}; rewirings described by $(d=0, r=0.5)$ were applied in Refs.~\cite{Kozma,Centola,Vazquez};
 those that can be classified as $(d=0, r=1)$ were utilized in Refs.~\cite{Durrett,BMin}; while Ref.~\cite{Kimura} used a fixed value of $d=0.5$ with varying $r$ and found fragmentation at $r=1$. Fragmentation occurs  regardless of the node dynamics
employed in these cases. Coevolution of any node dynamics with a rewiring process characterized by  $d=0$ or $r=1$  will eventually lead to fragmentation of the network, as the rate of application of this rewiring is increased.

\begin{figure}[h]
    \centering
    \includegraphics[scale=0.6]{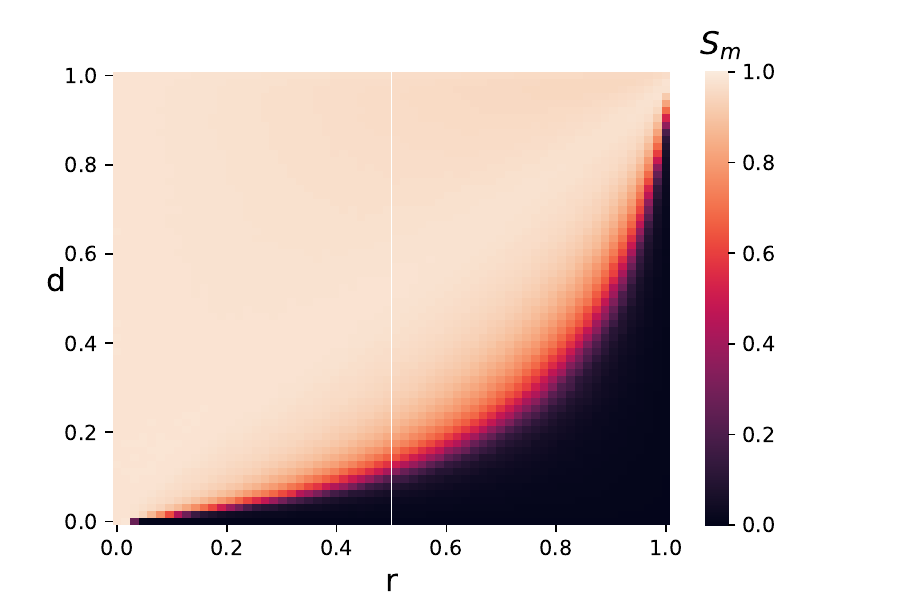}
    \caption{Average normalized size $S_m$ of the largest
    component on the plane $(r,d)$. Parameters are $N=3200$, $\langle k \rangle=4$,  $G=320$.
    Each data point shown corresponds to the average over $100$ realizations of initial random conditions for the network.}
    \label{fig:f3}
\end{figure}

In finite-size systems, however, fluctuations may eventually
drive the number of active links to zero  for parameter values in the vicinity of the boundaries $d=0$ or $r=1$ on the $(r,d)$ plane, leading to fragmentation of the network in numerical simulations.
To characterize the integrity of the network, we
employ the average normalized size (divided by $N$) of the largest component or subgraph in the system, regardless of the states of the nodes, denoted by $S_m$.
Figure~\ref{fig:f3} shows the quantity $S_m$ numerically calculated on the space of rewiring parameters $(d, r)$ for random initial networks. We distinguish two main regions on this plane: a region  for which $S_m = 1$, describing the presence of a large component; and a region  adjacent to the lines $d=0$ and $r=1$ where $S_m = 0$, consisting of small, separated components, corresponding to fragmentation of the network.

In addition to fragmentation, we investigate the emergence of modular or community structure in the network. We call a domain a set of connected nodes that share the same state. A community structure consists of the presence of several interconnected domains on a large component, where nodes are highly connected within a domain, with fewer connections between nodes in different domains.
In this case, the density of active links  $\rho$ is small but non-vanishing.
Conversely, a fragmented network is characterized by the presence of multiple disconnected domains in which no active links are present, yielding $\rho = 0$.

As a measure for the
emergence of modular structure, we define the modularity change $\Delta Q \equiv Q(t \to \infty) - Q(0)$, where
$Q(0)$ is the modularity of the initial random network and $Q(t \to \infty)$ is the stationary modularity of the network for
asymptotic times. Then, a value $\Delta Q = 0$
reflects the persistence of the initial random topology, while $\Delta Q > 0$ indicates an increase in modularity with
respect to the initial random network, independently of the absolute values of $Q$. We employ the Louvain community detection algorithm \cite{Louvain} to calculate the values of $Q$ in each case.

\begin{figure}[h]
    \centering
       \includegraphics[scale=0.6]{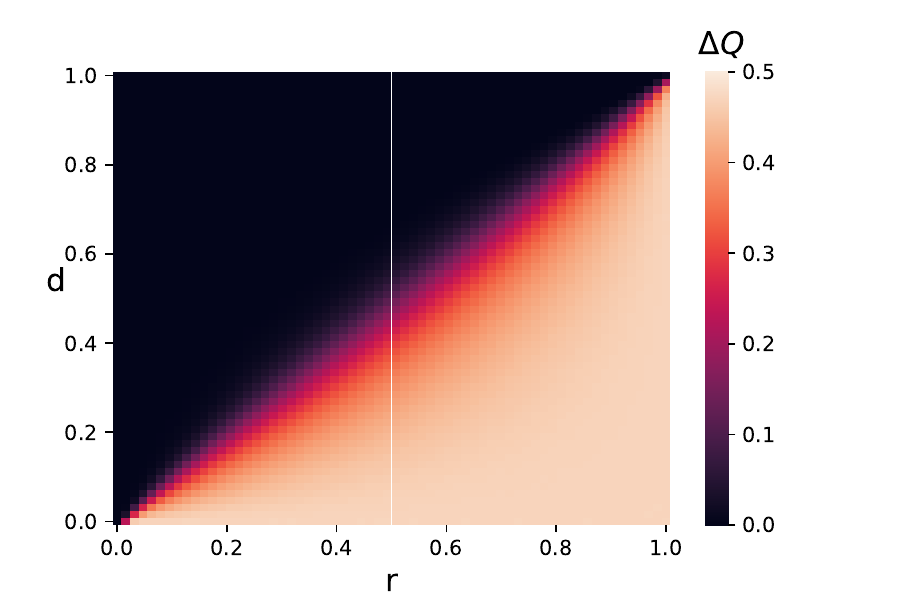}
\caption{Modularity change $\Delta Q$ numerically calculated on the plane $(d,r$). The values $\Delta Q$ are indicated by a color code on the right.
Parameters are $N=3200$, $\langle k \rangle=4$, $G=320$.
    Each value shown corresponds to the average over $100$ realizations of initial random conditions for the network.}
    \label{fig:f4}
\end{figure}

Figure~\ref{fig:f4}  displays the modularity change $\Delta Q$, averaged over several realizations of initial conditions for
random networks, on the space of parameters $(d, r)$. By comparing  Fig.~\ref{fig:f4} with Fig.~\ref{fig:f3}, we see that regions of parameters for
which $\Delta Q > 0$ include values $(d, r)$ where the network becomes fragmented. The Louvain algorithm considers a network
with separated subgraphs as modules, assigning maximum modularity to it. Thus, the quantity $\Delta Q$ alone cannot effectively
distinguish the existence of connected modules on a community from a fragmented network, since both situations yield $\Delta Q > 0$.

To identify the emergence of community structure, we define the product $\Delta Q \times S_m$ as an order parameter. If $\Delta Q \times S_m = 0$, then: i) either the modularity change vanishes ($\Delta Q = 0$), so that the network remains topologically equivalent to the initial random configuration; or ii) the size of the largest component is zero ($S_m = 0$), corresponding to fragmentation. In both cases, no community structure is present. Conversely, a positive value $\Delta Q \times S_m > 0$, reflects both a modularity increase and a nonvanishing large connected component, thus signaling the presence of a community structure.

\begin{figure}[h]
    \centering
    \includegraphics[scale=0.55]{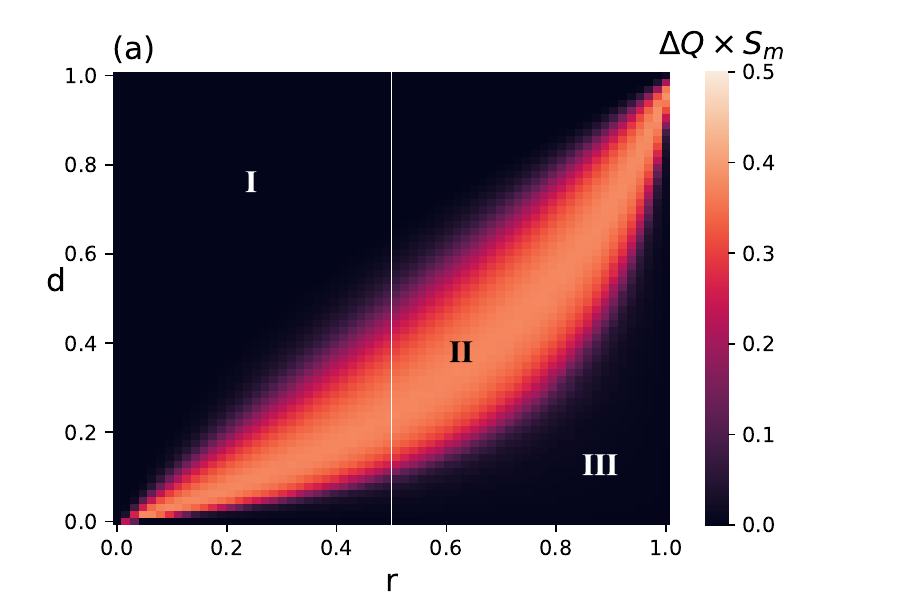}
        \includegraphics[scale=0.54]{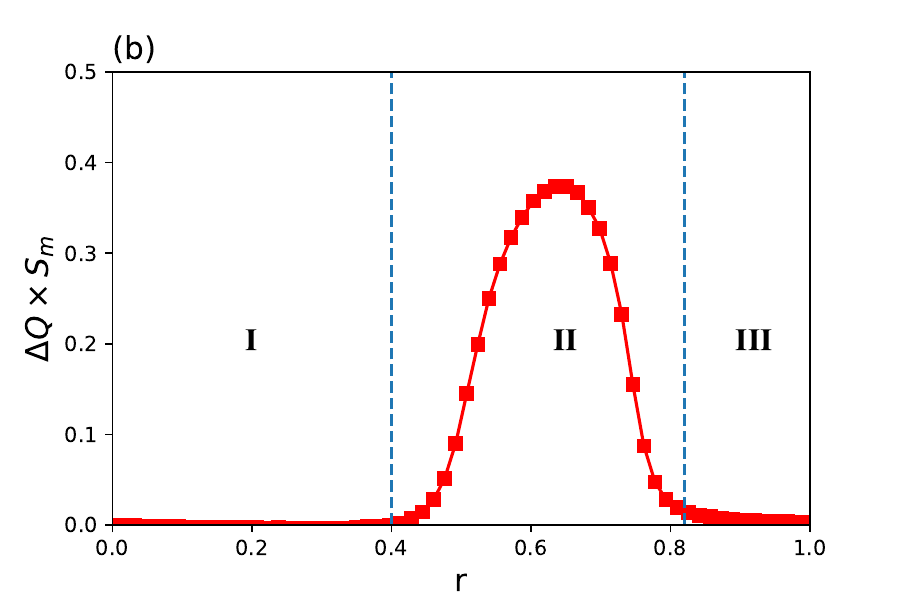}
    \caption[Quantity $\Delta Q \times S_m$ calculated on the space of parameters $(d,r)$.]{(a) Quantity $\Delta Q \times S_m$ calculated on the space of parameters $(d,r)$. The values of $\Delta Q \times S_m$ are indicated by a color code bar on the right. (b) $\Delta Q \times S_m$ as a function of $r$ for parameters along the diagonal $d=1-r$.
Network parameters are: $N=3200$,  $\langle k \rangle=4$, $G=320$.  On both panels, each data point corresponds to the average over $100$ realizations of initial random conditions for the network.}
    \label{fig:f6}
\end{figure}

Figure~\ref{fig:f6}(a) shows the quantity $\Delta Q \times S_m$ calculated on the space of the parameters $(d, r)$.
Three regions, associated with different structures or ``phases'' of the network, can be distinguished on the
plane $(d, r)$ in Fig.~\ref{fig:f6}(a): (I) a region where $\Delta Q \times S_m = 0$ corresponding to a large random subgraph, characterized by
$\Delta Q = 0$, $S_m = 1$; (II) a region having $\Delta Q \times S_m >0$
where community structure emerges on a connected graph, corresponding to $\Delta Q > 0$, $S_m = 1$; and (III) a region where the network is fragmented in small separated components; characterized by $\Delta Q > 0$, $S_m = 0$, and the product $\Delta Q \times S_m = 0$.

Figure~\ref{fig:f6}(b) shows
the product $\Delta Q \times S_m$ as a function of $r$ along the diagonal line
$d = 1-r$.
Community structure appears for intermediate values of the parameters $r$ and $d$ on region II. Progressing along the line $d = 1 - r$ describes a continuous transition from the point $(d, r) = (1, 0)$, representing complete heterophily or tolerance, to the point $(d, r) = (0, 1)$, representing complete homophily or intolerance. In the vicinity of these limits, $\Delta Q \times S_m \approx 0$, indicating that no community structure can be sustained under either extreme heterophily or extreme homophily. Instead, Fig.~\ref{fig:f6}(b) shows that communities arise only at intermediate values of the rewiring parameters $d$ and $r$, corresponding to regimes of moderate tolerance.

The decay of active links leads to an increase in the  modularity of
the network. As communities emerge, only a small fraction of active
links persist between nodes belonging to different domains, while the
majority become inert, connecting nodes that share the same state within a domain. A large fraction of inert links indicates a predominance of homophilic interactions, which is a condition for many community detection algorithms, such as the Louvain method, to identify densely intra-connected modules and thereby yield high modularity. Consequently, community structure can be characterized by the coexistence of a giant connected component ($S_m \approx 1$) and a high density of inert links $(1 - \rho)$. The product $(1 - \rho) \times S_m \to 1$ thus provides a quantitative measure of community formation in the network.

\begin{figure}[h]
    \centering
\includegraphics[scale=0.6]{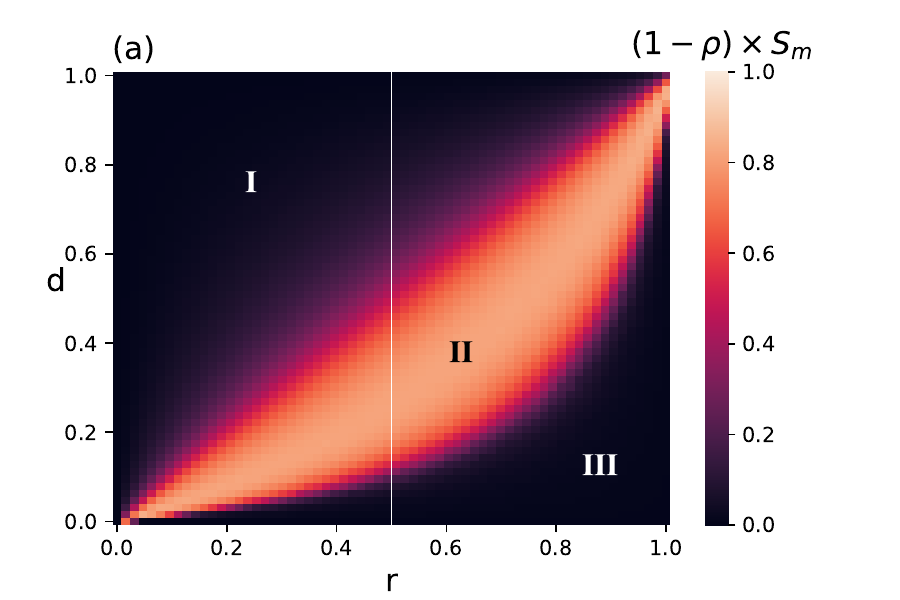}
\caption{Product $S_m \times (1 - \rho)$ on the plane $(d, r)$. The values of $S_m \times (1 - \rho)$ are indicated by a color code
bar on the right.
Region labeled II, where $S_m \times (1 - \rho) \to 1$, corresponds to the emergence of community structure. In regions I and III,  $S_m \times (1 - \rho) \approx 0$.
Network parameters: $N = 3200$, $\langle k \rangle= 4$, $G = 320$. Each point shown corresponds to the average over $100$ realizations
of initial random conditions for the network.}
    \label{fig:fP}
\end{figure}

Figure~\ref{fig:fP} shows
the quantity $(1-\rho) \times S_m$ calculated on the space of parameters $(d,r)$.
Three regions are labeled in Fig.~\ref{fig:fP}.
In the ``half moon'' region labeled II, the product
$(1-\rho) \times S_m \to 1$, indicating the presence of community structure; while  $(1-\rho) \times S_m  \to 0$ in both regions labeled I and III.
 Region II in Fig.~\ref{fig:fP} approximately coincides with the community phase region II in Fig.~\ref{fig:f6}.
 Both figures reveal that
 the formation of
communities requires the presence of more homophily (that is, $r > d$) than heterophily in the interactions between agents, but not extreme homophily that would lead to fragmentation of the network.

We have employed an initial uniform distribution of node states ($N/G=$ constant) to maximize the probability of interaction between diverse groups and to maintain the mean-field assumption of homogeneity. For non-uniform initial distributions, we expect the qualitative aspect of the phase diagram in Figures~\ref{fig:f6} and \ref{fig:fP} -- specifically the existence of the three distinct network phases -- to remain robust, only with shifted boundaries.

Note that our choice of the ratio $N/G=10$ guarantees the emergence of a dense community structure. In the limit $N/G \to 1$, where very few nodes share the same state, the community structure phase II would effectively vanish. In this situation, the mechanism for homophilic clustering fails; nodes cannot find similar others to form a module. Consequently, the system would have a random network phase and a fragmented phase for very large times, without sustaining the robust community structure observed at higher ratios. In the case $N/G=1$, only a random network phase exists.

\begin{figure}[h]
    \centering
    \includegraphics[scale=0.28]{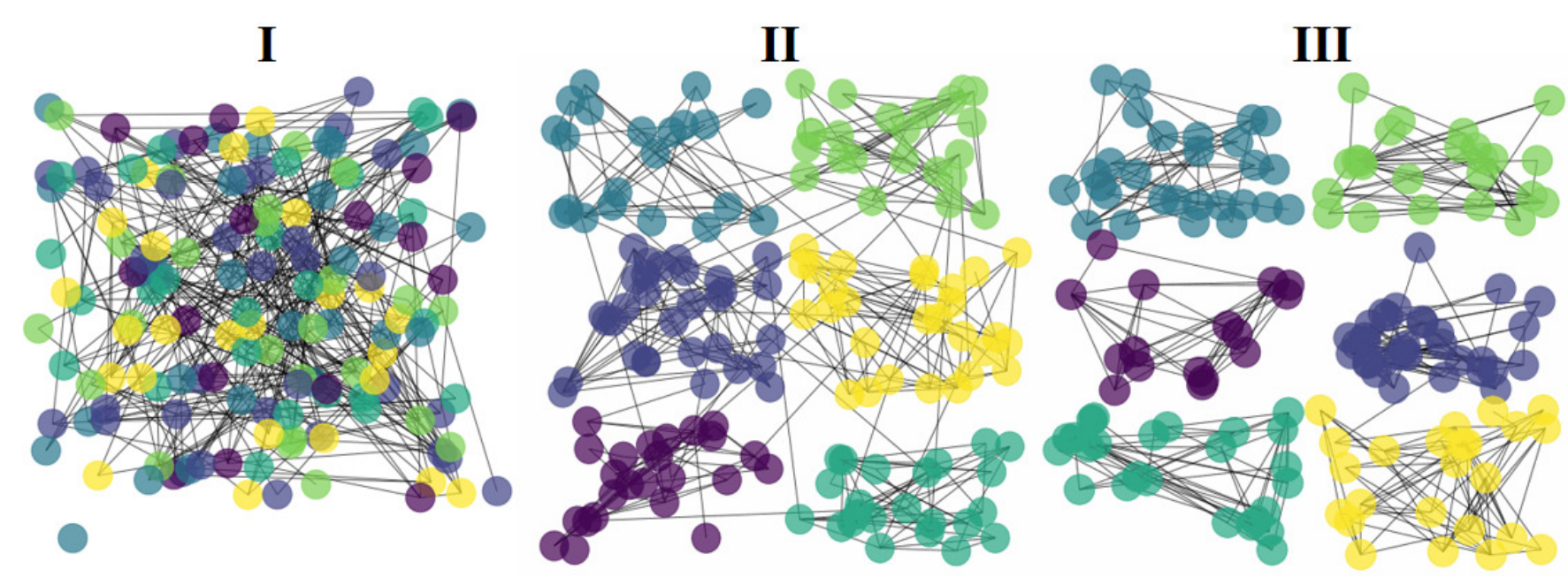}
    \caption{Network configurations corresponding to the regions labeled I, II, and III, on the space of parameters $(d,r)$ in Fig.~\ref{fig:f6}.  (I) connected random network; (II) community structure; (III) fragmented network. Network parameters are: $N=150$,  $\langle k \rangle=4$, $G=6$. The node states are indicated by different colors.}
    \label{fig:f7}
\end{figure}

Figure~\ref{fig:f7} illustrates the asymptotic network configurations obtained from numerical simulations for parameter values in regions labeled I (connected random network), II (community structure), and III (fragmented network) on the plane $(d, r)$ of Fig.~\ref{fig:f6}.

\section{Conclusions}
We have investigated the effects of adaptive rewiring
in networks with fixed node states, where only the links evolve.
This configuration represents a limiting case of coevolutionary dynamics in which the time scale of node state changes is much slower than that of the rewiring process.
We employed the general framework for coevolutionary dynamics where a rewiring process is parametrized through two parameters $d$ and $r$, which denote the probabilities of disconnection and connection between nodes in identical states, respectively  \cite{Herrera2011}. The space of parameters
 $(d, r)$ thus provides a continuous representation of the homophilic and heterophilic tendencies in the system, enabling a systematic analysis of how adaptive rewiring influence network topology.

We derived a mean-field solution for the density of active links  that connect nodes in different states.
This formulation provides an explicit stationary solution $\rho^*$, which is in good agreement with numerical simulations, particularly in the symmetric regime and in the homophilic and heterophilic limits.
We found that, in the mean-field limit, the network undergoes fragmentation when $d = 0$ or $r = 1$, corresponding to vanishing density of active links. For finite-size systems, fluctuations near these parameter values can also drive the network into fragmentation.
Our non-trivial observation that network fragmentation is independent of the node dynamics employed and just a consequence of extreme homophilic rewiring processes  contributes to simplify the current understanding of coevolutionary dynamics.

To distinguish the emergence of communities from simple fragmentation, we introduced the order parameter $\Delta Q \times S_m$, combining modularity increase and the size of the largest connected component in the network. This criterion effectively identifies three distinct phases on the space of parameters $(d,r)$: (I) random connected network, (II) community structure, and (III) fragmented network.
Additionally, we calculated the product $(1-\rho)\times S_m$, which characterizes community formation as the coexistence of a giant component with a
low density of active links.
Community structure emerges only for intermediate values of the rewiring parameters, specifically under moderate homophily ($r > d$). It disappears under extreme heterophily or homophily, which instead drive the network toward randomness or fragmentation, respectively.
The existence and boundaries of the community structure phase are key findings that provide a deeper understanding of how adaptive rewiring induces self-organization before the transition to fragmentation.

These results show that adaptive rewiring
can induce structural transitions in networks,
including the emergence of community structure and fragmentation.
The mean-field approach yields predictive insights into these transitions, while the proposed order parameters provide robust criteria for identifying community formation.
In particular, characterizing the transition to network fragmentation
has implications for current research on the formation of isolated groups and echo chambers, where individuals are primarily exposed to their own opinions.
Our results are relevant to a wide range of systems in which node states are static or change slowly, including ideological social networks, identity-based social segregation, technological compatibility networks, and ecological systems.

We point out the analogy between our model and the   continuous swarmalator model \cite{Okeefe,Yadav} in the limit $K=0$, where the phases or states of the agents are static and their spatial positions evolve. The spatial distance becomes inversely related to interaction strength -- effectively similar to adaptive rewiring, where the link probability depends on state similarity. Both systems share a common underlying mechanism: the minimization of state/phase difference through structural rearrangement -- be it in physical space or network topology.
Consequently, the self-organized states in both scenarios appear similar.

Future extensions for investigation include the generalization of the rewiring process for continuous and for multivariable node states; studying the topological effects of
preferential attachment rules for the connection
action, and the consideration of variable connection strengths.

\section*{Acknowledgment}
This work was supported by ViceRectorado de Investigaci\'on and Innovaci\'on, Universidad Yachay Tech, Ecuador.

\end{document}